\documentclass[fleqn,10pt]{wlscirep}
\usepackage[utf8]{inputenc}
\usepackage[T1]{fontenc}
\usepackage[noabbrev]{cleveref}
\usepackage[separate-uncertainty=false]{siunitx}

\graphicspath{{figures/}}

\title{Directional optical parametric amplification in a hyperbolic metamaterial}

\author[1]{Antonio Gianfrate}
\author[1,*]{Dimitrios Trypogeorgos}
\author[1]{Paolo Comaron}
\author[2]{Dawid Paszko}
\author[2]{Marzena Szyma\'nska}
\author[1]{Milena De Giorgi}
\author[1]{Dario Ballarini}
\author[1]{Daniele Sanvitto}
\affil[1]{CNR Nanotec, Institute of Nanotechnology, via Monteroni, 73100, Lecce, Italy}
\affil[2]{Department of Physics and Astronomy, University College London, Gower Street, London, WC1E 6BT, United Kingdom}

\affil[*]{dimitrios.trypogeorgos@cnr.it}


\begin{abstract}
Optical parametric amplification (OPA) comprises essentially a nonlinear four-wave mixing process in which a `pump' and a `signal' field give rise to an `idler' field under certain phase-matching conditions.
Here we use a photonic crystal waveguide strongly-coupled with an excitonic reservoir to generate this process between different guided modes at optical wavelengths.
Differently from classical nonlinear optical crystals, where the pump and idler photons travel almost collinearly, our exciton-polaritons are naturally separated in the waveguide due to their opposite group velocities.
Due to the high efficiency of the process we can generate the idler field of the parametric process by pumping with a continuous wave laser and choose its direction of propagation in the waveguide by adjusting the angle of incidence of the seed laser.
We show the OPA process to be robust against surface defects of the waveguide and can lead to simple-to-fabricate devices compared to microcavities that take advantage of strong signal-idler correlations in a propagating geometry.
Our results closely agree with mean-field numerical simulations.
\end{abstract}

\begin{document}

\flushbottom
\maketitle

\thispagestyle{empty}

\section*{Introduction}

The nature of nonequilibrium stationary states of light-matter quasiparticles in GaAs nanostructures is strongly shaped by the way the system is driven by the laser~\cite{RevModPhys.85.299}. 
When the laser frequency is tuned to resonantly excite highly energetic states, the resulting particles undergo intricate scattering processes before eventually condensing into a polariton branch. 
These scattering events erase any coherence initially provided by the laser, preventing it from being transferred to the polaritonic states. 
By contrast, in a coherent driving scheme, particles are directly excited by a monochromatic laser, resonant with or near the dispersion~\cite{ciuti.PhysRevB.63.041303}. 
This setup bypasses the complex scattering processes, enabling a first-principles description rather than relying on phenomenological approximations~\cite{PhysRevLett.93.166401,whittaker.PhysRevB.63.193305,whittaker.PhysRevB.71.115301,wouters.PhysRevB.75.075332,PhysRevA.76.043807}. 
With coherent driving, two symmetries play a key role in the dynamics of the system~\cite{ciuti.PhysRevB.63.041303}: $U(1)$ symmetry, associated with phase rotations of the polariton field, and spatial rotational symmetry.

There exist a pumping scheme of particular interest, when the system is driven by a coherent laser pump with energy $\omega_p$ and momentum $k_p$, which can scatter in pairs to other available states while conserving both energy and momentum; this phenomenon characterizes the optical parametric oscillator (OPO) regime~\cite{stevenson.PhysRevLett.85.3680,ciuti.PhysRevB.63.041303}. 
Once the pump strength surpasses a certain threshold, parametric scattering populates two additional states, the signal and idler, with respective energies $\omega_s$, $\omega_i$ and momenta $k_s$, $k_i$, which fulfil the conditions $2\omega_p = \omega_s + \omega_i$ and $2\mathbf k_p = \mathbf k_s + \mathbf k_i$~\cite{whittaker.PhysRevB.63.193305}. 
The intense and directional emission from these modes~\cite{baumberg.PhysRevB.62.R16247} holds promise for optical applications~\cite{Gavrilov2007,PhysRevLett.101.136401}. 
Furthermore, coherently pumped polariton systems have become a valuable framework for investigating collective many-body effects.
Recent advances in semiconductor fabrication and the large $\chi_3$ values in exciton-polaritons enabled the achievement of micrometric-size footprint and low-threshold OPOs~\cite{stevenson.PhysRevLett.85.3680,ciuti.PhysRevB.63.041303,whittaker.PhysRevB.63.193305,baumberg.PhysRevB.62.R16247,saba.nature2001,messin.PhysRevLett.87.127403}.

When using planar microcavities, to satisfy the phase matching condition, the external pump is commonly set to the inflection point of the lower polariton branch. 
This results in a non-degenerate signal and idler components of highly different excitonic fractions, affecting both the lifetime and the group velocity of these signals. 
While such OPO processes have been useful for investigating the generation and dynamic of topological defects in the polariton quantum fluid, this asymmetry between signal and idler has proven less suitable for studying, and in turn, exploiting the spatio-temporal correlations between the generated signals.

To overcome this limitation, while still satisfying the phase-matching condition, a natural approach has been identified in band engineering~\cite{dasbach.PhysRevB.66.201201,ardizzone.scientific.reports2013,trypogeorgosEmergingSupersolidityPolariton2024,diederichs.nature.2006,abbarchi.PhysRevB.83.201310}.
Over the years, several different geometries have been explored, including double and triple cavities, wires, and pillars.
The aim has been to ``tilt'' the OPO process to achieve isoenergetic and isomomentum processes of signal and idler generation.
In such process $\omega_s = \omega_i$, and for a normally incident pump $\mathbf k_p =0$ the signal and idler are counterpropagating ~\cite{dunnett.PhysRevB.98.165307,spanoCoherencePropertiesExciton2012} and can be separately collected outside the pumping spot region. 
However, attempts to engineer such processes using microcavity variants inevitably lead to inefficient photon extraction and short propagation distances of the signal and idler beams.

Here instead, we employ a slab waveguide geometry naturally able to confine the generated signals by total internal reflection and consequently suitable for manufacturing devices based on propagating photons; the signal and idler beams are generated on counter-propagating modes and can be selectively addressed.
The dispersion engineering process is done by designing a photonic crystal grating, through which it becomes possible to selectively position counter-propagating modes and tune their mutual Rabi coupling. 
The introduction of a unidimensional grating on top of a planar slab results in highly anisotropic dispersion with hyperbolic geometry, leading to pronounced directionality~\cite{georgakilas_quantum_2024}.
Recently, on to this platform, we demonstrated the non resonantly driven onset of an isoenergetic OPO process taking place between the BiC condensate lying at the $\Gamma$ point and a couple of symmetric higher order modes~\cite{ardizzonePolaritonBoseEinstein2022,efthymiou-tsironiCondensationDynamicsTwodimensional2024,gianfrateReconfigurableQuantumFluid2024,nigroSupersolidityPolaritonCondensates2024,riminucciPolaritonCondensationGapConfined2023,trypogeorgosEmergingSupersolidityPolariton2024}. 
In this work we demonstrate how is is possible to trigger this effect in a resonant fashion using the lossy branch of the Dirac cone. 
Using a seed beam we force the optical parametric amplification of the idler beam and can chose its angle of propagation at will.

\section*{Results}

After a brief reminder of the physics of the OPO from a theoretical perspective, we describe the one-dimensional energy-momentum dispersion of our photonic crystal waveguide and its relevant topological properties.
We then extend this description to the complete two-dimensional reciprocal space at the energy manifolds of interest.
The phase-matching condition for the OPA takes place in this space at the energy of the lossy branch.
By switching the pump and seed signals on and off and analysing the photolumiscence as a function of the wavevector, we demonstrate OPA at various angles and show how this process is robust against positional variations on the sample.

\subsection*{Theory background}

The OPO regime is often analyzed theoretically using a simplified three-mode model, where the mean field includes only the dominant signal, pump, and idler modes~\cite{https://doi.org/10.1002/pssb.200560961,C_Ciuti_2003,wouters.PhysRevB.75.075332,PhysRevB.93.195306}.
Using the simplest assumption of three modes, $\psi = \psi_s + \psi_p + \psi_i$, representing the signal, pump, and idler respectively, and treating each mode as a plane wave, one can derive a three-mode framework for describing the polariton OPO regime, consisting of three coupled complex Gross-Pitaevskii equations~\cite{whittaker_polariton_2017,dunnett.PhysRevB.98.165307}.
The resulting set of three complex equations is invariant under a global
U(1) phase rotation, 
$S \to S' = Se^{i\alpha}, \; I \to I' = Ie^{-i\alpha}, \; P \to P' = P.$
The relative phase of the signal and idler states is free, and, consistent with standard driven-dissipative condensates, the U(1) symmetry can be broken spontaneously~\cite{RevModPhys.85.299} in each realisation.
This feature is responsible for the appearance of a gapless Goldstone mode above the OPO threshold~\cite{PhysRevB.74.245316, ballarini_observation_2009}.
As such, universal critical phenomena such as long-distance coherence~\cite{PhysRevB.74.245316}, superfluidity~\cite{PhysRevB.92.035307}, conventional~\cite{PhysRevX.5.041028,PhysRevResearch.5.043286} and unconventional~\cite{PhysRevLett.130.136001} topological order can emerge. 

The three-mode ansatz framework, however, suffers from an important deficiency~\cite{dunnett.PhysRevB.98.165307}. 
This description provides insufficient constraints to uniquely identify the signal and idler momenta. 
Instead, in experiments and numerical simulations of the complete problem, there are no limits on the number of modes that can be occupied; the system naturally selects a unique momentum configuration, which is typically dominated by one specific $k_s, k_i$ pair~\cite{whittaker_polariton_2017}.
A more complete description of the system that considers solutions beyond the three-mode ansatz is given by the numerical solution of the multimode polariton field integrated in the full real space~\cite{ciuti.PhysRevB.63.041303,wouters.PhysRevB.75.075332}.

A widely used model, which addresses the aforementioned problems, describes the polariton field using a stochastic complex Gross-Pitaevskii equation (scGPE). 
This model is derived by mapping the time evolution of the quasiprobability function from the Fokker-Planck equation onto a Langevin equation.
We note that the complete stochastic version of the aforementioned model can be obtained in the semiclassical approximation of Keldysh field theory~\cite{PhysRevB.93.195306}. 
This allows for the incorporation of effects at all levels in classical fields and up to the second order in quantum fluctuations \cite{PhysRevB.89.134310}, making this approach nearly precise for systems containing a large number of particles.

Through nonlinear stochastic simulations, it has been found that OPO exists for a range of pump strengths between a lower and upper threshold and its behaviour is categorized into four regimes~\cite{dunnett.PhysRevB.98.165307}: three-mode solutions near the upper threshold, multimode solutions with significant satellites at higher intermediate pump strengths, approximate three-mode solutions at lower intermediate pump strengths with smaller satellites, and ring patterns near the lower threshold.
Comparison with the full stochastic modeling for OPO polaritons 
beyond the mean-field approximation~\cite{dunnett.PhysRevB.98.165307} reveals the regimes where the commonly used three-mode OPO description is valid.
The three-mode ansatz effectively describes the system in regimes around the upper threshold and at lower intermediate pump strengths, but not at higher intermediate pump strengths and at the vicinity of the lower threshold, where multimode effects dominate.

Additionally, in the three-mode ansatz framework, signal momentum can be predicted through a basic stability analysis, eliminating the necessity to solve the complete set of nonlinear multimode equations: linear-response analysis of the pump-only system provides reliable predictions for the signal momentum near the upper threshold and the size of the rings near the lower threshold. 
However, its accuracy diminishes for intermediate pump strengths. 
Also, the signal momentum remains largely independent of the pump laser detuning and varies minimally with pump momentum, while the signal energy increases with pump energy and momentum, aligning with previous experimental observations.

It is worth noting the type of order that forms when the OPO thresholds are crossed.
Assuming constant mean-field amplitudes,
the OPO regime exhibits density-wave and time-crystal order, characterized by a base wave vector \( k_{si} = 2 \) and a frequency \( \omega_{si} = 2 \), respectively \cite{PhysRevX.7.041006}. 
Here the Goldstone mode corresponds to fluctuations in the relative phase between signal and idler.
In possessing a gapless Goldstone mode, such a system is expected to be a superfluid according to the traditional definitions. 
A combined theoretical and experimental study revealed that the coupling between the three OPO modes gives rise to complex nonlinear behaviour~\cite{PhysRevB.92.035307}.

\subsection*{Dispersion}

Our working device is a GaAs photonic crystal waveguide in the strong coupling regime (see Methods).
The waveguide supports propagating modes, outside the light cone, at optical near-infrared wavelengths.
The propagating modes are also strongly coupled to the exciton resonance which is largely non-dispersing for the energies and wavevectors considered here.
A one-dimensional, sub-wavelength etched grating folds and couples the TE$_0$ and TE$_1$ counter-propagating modes.
By tuning the period of the grating, \qty{242}{nm}, we arranged for the diffractive mode coupling for the TE$_0$ modes to occur in the vicinity of the exciton resonance at \qty{1.527}{eV}.

The complex nature of the diffractive coupling leads to a non-trivial topology and a redistribution of losses between the diagonalized branches, \cref{eq.H}. 
Engineering the phase of the coupling makes it possible to achieve a state with vanishing linewidth, the BiC state, alongside a counterpart branch characterized by twice the losses, the lossy state. 
While the vanishing linewidth is a clear advantage for condensation, as it contributes to lowering the threshold~\cite{riminucci_polariton_2023,ardizzonePolaritonBoseEinstein2022}, it also inhibits for resonant injection. 
The lossy state on the other hand, lying just a few meV above the BiC, provides incredibly efficient access. 
The M-shaped geometry of that dispersion branch fulfils the phase-matching condition for different injection angles, however we concentrate on the high-symmetry $\mathbf k=0$ point that leads to energy-degenerate OPO and OPA.

\begin{figure}
    \centering
    \includegraphics[]{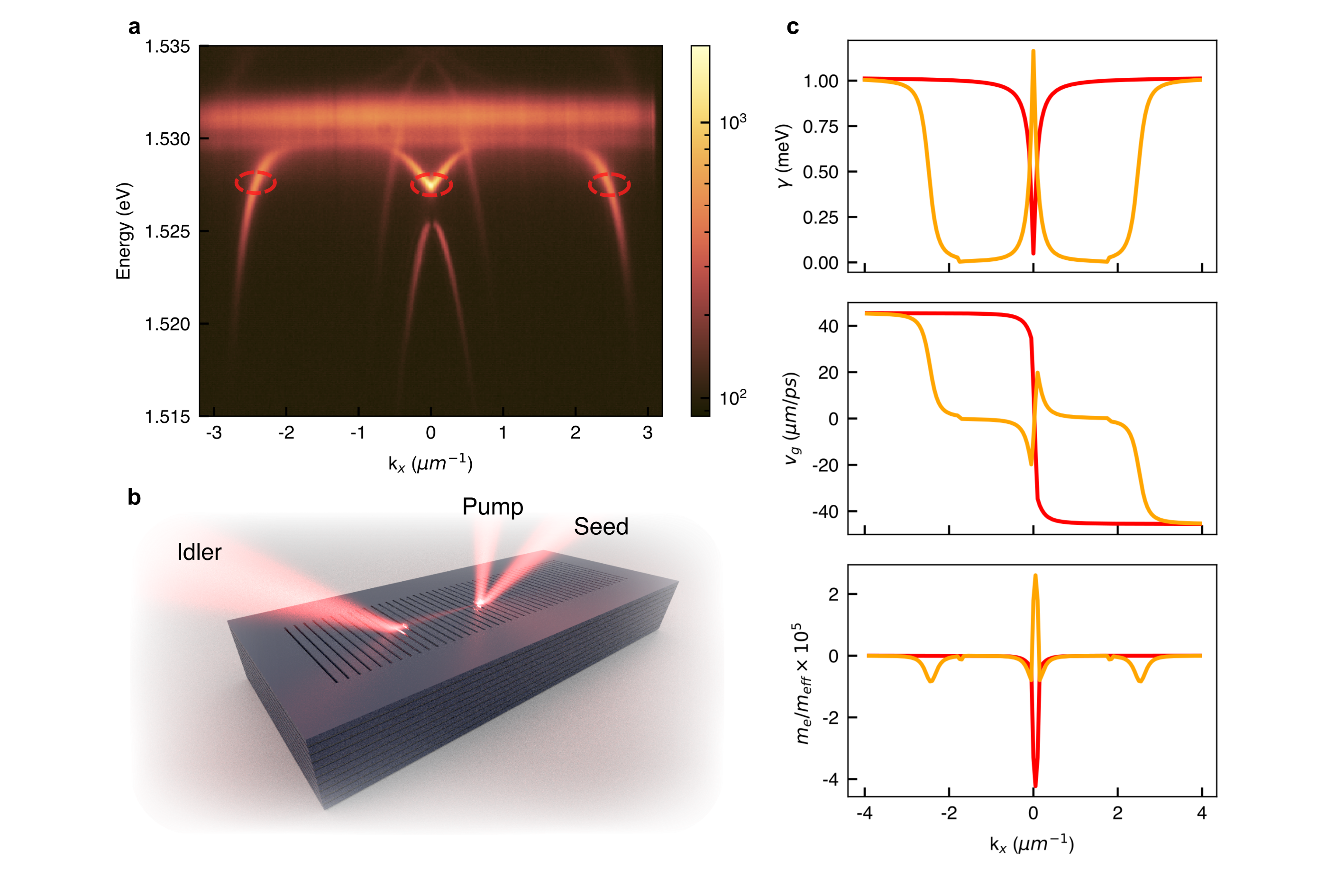}
    \caption{\textbf{System dispersion characteristics.} \textbf{a.} The energy-momentum dispersion of the waveguide shown in \textbf{b.} along $k_x$. The M-shaped lossy branch is excited resonantly with a pump and parametrically scatters to the TE$_1$ modes. The idler is represented as an in-plane propagating wave. \textbf{c.} The linewidth, group velocity, and inverse effective mass of the lowest two energy branches.}
    \label{fig:1}
\end{figure}

To characterize the system dispersion we used \qty{800}{nm} nonresonant light to excite the sample at approximately \qty{20}{meV} detuned above the exciton resonance. 
The resulting far field, energy-resolved pattern along $k_x$ is shown in \cref{fig:1}a.
Along this direction we can clearly distinguish the two counter-propagating TE$_0$ modes anti-crossing at the $\Gamma$ point, hosting the BiC on to the negative mass branch and the lossy on to the positive one.
The losses vanish at the BiC branch and become twice as much at the lossy, as can be seen directly by the linewidth of the photoluminescence, \qty{1.25}{m eV} at $
\mathbf k=0$, in \cref{fig:1}a and the imaginary part of the diagonalised eigenergies of \cref{eq.H} in \cref{fig:1}b.
At the energy of the lossy and wavevectors roughly $k_x=\qty{\pm2.45}{\mu m^{-1}}$ lie a pair of TE$_1$ modes that are propagating with a large group velocity of \qty{45}{\mu m/ps}, see \cref{fig:1}c.

\Cref{fig:1}d shows the effective mass of the M-branch along $k_x$ that changes sign at the points where the TE$_1$ modes are coupled to the exciton.
At the $\Gamma$ point the mass is positive and equal to $2.8\times 10^{-6} m_e$, where $m_e$ is the mass of a free electron, c.f. the typical polariton mass in microcavities which is 2 orders of magnitude larger.
The small mass and effective detuning from the exciton leads to the lossy branch being strongly coupled with the exciton at $\pm \qty{0.2}{\mu m^{-1}}$.
This is comparable to the typical reciprocal waist of our resonantly injected beam, leading to a quasi-resonant condition at almost all detunings between the lossy and the exciton energy.

\subsection*{Two-dimensional reciprocal space}

Most of the dispersion characteristics can be obtained from the one-dimensional picture above, however we seed the OPA in iso-energetic reciprocal planes at the energy of the lossy.
The far-field picture of such planes can be adequately captured via a simple geometric construction. 
Each mode corresponds to a propagating cone with a finite aperture, folded by the diffractive grating within the light cone. 
At degeneracy, i.e., along a concave parabola in the $k_x = 0$ direction, the diffractive coupling induces a gap, resulting in a high aspect ratio, saddle-shaped lower branch with open, hyperbolic isofrequency curves and a mass ratio of about 100 along the principal reciprocal axes~\cite{georgakilas_quantum_2024,Drachev:13:Optica}.
The upper branch on the other hand shows ellipsoidal isofrequency curves. 

\begin{figure}
    \centering
    \includegraphics[]{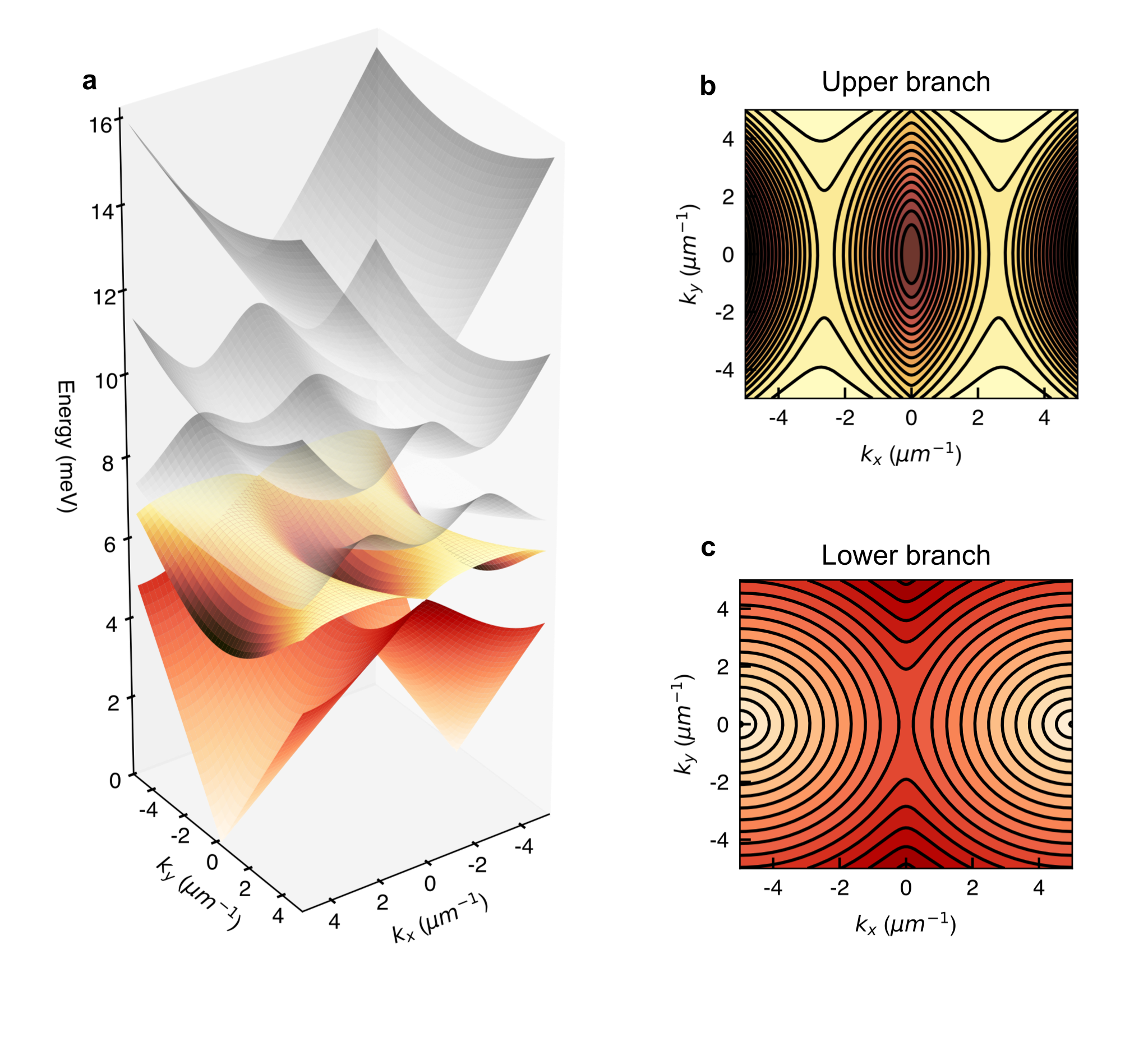}
    \caption{\textbf{Two-dimensional reciprocal space.} \textbf{a.} Eigenvalues of the system hamiltonian, see Methods for details, showing five branches appearing below and above the exciton energy at \qty{7}{meV} in these reference units. The iso-energetic surfaces at the energies of the lossy are ellipsoids,\textbf{b}, while at the energy of the BiC are hyperboloids, \textbf{c.}}
    \label{fig:2}
\end{figure}

Another set of cones, further displaced from the fundamental ones by approximately \qty{2.45}{\mu m^{-1}}, leads to the TE$_1$ coupling. 
An iso-energetic surface of the dispersion at the lossy is consequently composed by a strongly elongated ellipsoid at the $\Gamma$ point paired with a couple convex modes with a radius of curvature \qty{13}{\mu m^{-1}}, see \cref{fig:2}.

\subsection*{Directional phase-matching}

We used a continuous-wave Ti:sapphire laser to resonantly excite polaritons at the $\mathbf k=0$ point of our photonic crystal waveguide. 
The laser energy was set to \qty{1.5276}{eV}, resonant with the lossy branch.
The pump waist in reciprocal space is estimated to be \qty{0.1}{\mu m^{-1}} (FWHM) and in coordinate space is \qty{60}{\mu m} for both the pump and seed beams. 
This choice of waist is driven, on one hand, by the necessity of excitation selectivity and, on the other hand, by the need to avoid exceeding the lateral dimension of the \qty{50}{\mu m} grating. 
However, despite the small reciprocal-space waist of the excitation spot, the combination of the small polaritonic mass along the $k_x$ direction~\cite{georgakilas_quantum_2024} and the high group velocity of the bare photonic modes results in an excitation scheme with a pump unconditionally quasi-resonant.
Additionally, the small curvature of the confined modes that compose the dispersion, relative to the mode waist in reciprocal space, ensures that the phase-matching condition is fulfilled over a substantial portion of the TE$_1$ modes, ultimately resulting in a system with poor mode selectivity. 
For the same reason it was not possible to observe optical limiter behaviour~\cite{baas.PhysRevA.69.023809,bajoni.PhysRevLett.101.266402}.

\begin{figure}
    \centering
    \includegraphics[]{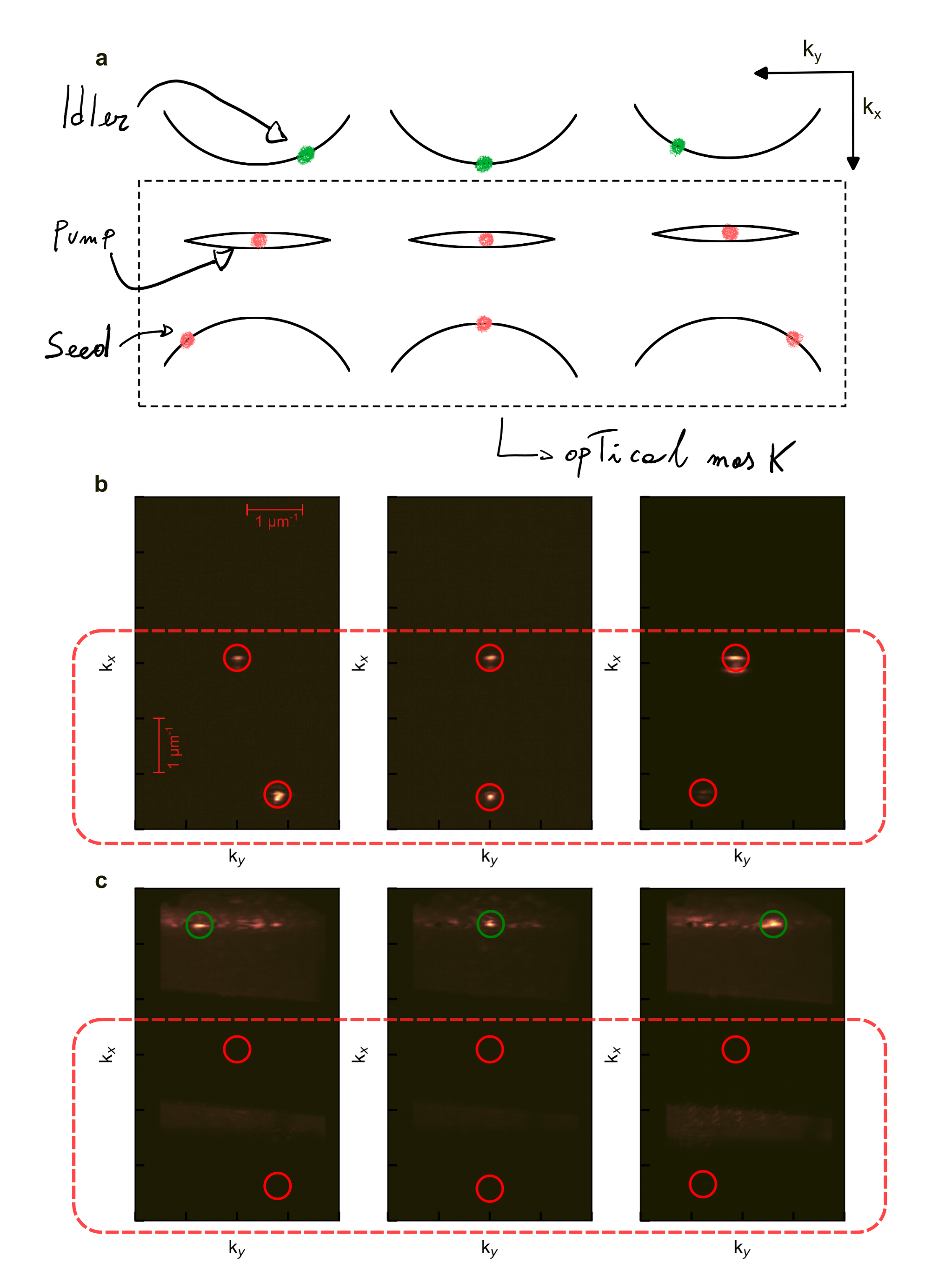}
    \caption{\textbf{Phase matching.} \textbf{a.} Schematic representing the idler beam generated by the iso-energetic OPA process changing angle to mirror the position of the seed beam. \textbf{b.} Three different positions of the seed laser with respect to the pump which is always at $k=0$. With the pump and seed beams masked \textbf{c.} the PL from the opposite TE$_1$ mode can be collected and it shows the conjugate generation of the idler beam.}
    \label{fig:3}
\end{figure}

The introduction of a seed beam resonant with the TE$_1$ mode breaks the symmetry along $k_x$, restoring mode selectivity and producing a clearly localized parametrically amplified signal. 
\Cref{fig:3} shows three different excitation configurations in which displacing the seed beam, resonant with the TE$_1$ branch, along the $k_y$ direction results in a signal centro-symmetrically displaced with respect to the $\Gamma$-point, unequivocally demonstrating that the phase matching condition is fulfilled.
Notice that in the left and right columns, Rayleigh scattering also increases at the $k_y$ wavevector of the seed but at the idler branch. 
This non-uniform scattering is due to the waveguide patterning that makes the two-dimensional reciprocal space non-uniform but induces a preferential direction along the one-dimensional grating. 
Along this direction, $k_x$ the grooves of the grating seem to Rayleigh scatter more strongly than the orthogonal direction, $k_y$.

\subsection*{Power dependence}

\begin{figure}
    \centering
    \includegraphics[]{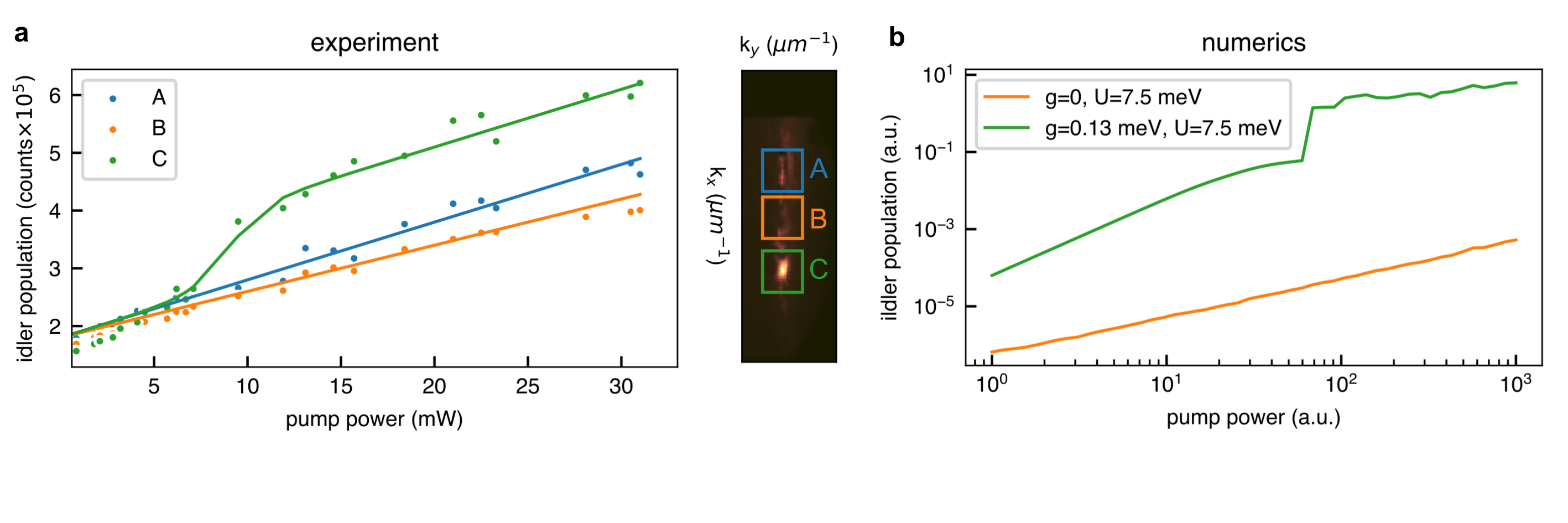}
    \caption{\textbf{OPA branch PL versus pump power.} \textbf{a.} The three curves represent the integrated PL collected from the ROIs depicted on the right. 
    Region C is phase-matched and shows distinct nonlinear behaviour. \textbf{b.} Linear and nonlinear behaviour is emulated in numerics by setting $g=0$ and $g\neq 0$. The numerical data cover a larger power range than the experimental ones.}
    \label{fig:4}
\end{figure}

\Cref{fig:4} shows the dependence of the idler beam on the power of the pump.
For these measurements we used the right-most configuration of \cref{fig:3} where the seed is resonant at $(k_x,k_y)=(-0.8,-2.4)$ and its intensity is kept constant at \qty{89}{W/cm^2}.
The photoluminescence at the idler branch can be divided into three regions of interest (ROIs) A, B, and C, that are distinguished by their $k_y$ value equal to \qtylist{-0.8;0;0.8}{\mu m^{-1}}.
These correspond respectively to the direct linear scattering of the seed, the linear scattering of the pump, and the parametrically generated signal due to the OPA.

\Cref{fig:4} shows the dependence of the PL collected in the three regions as a function of pump power when the seed laser is present.
Given the position of the seed in reciprocal space it is not possible for it to be phase-matched by itself and instigate any non-linear process.
As such the PL collected in region A, although it is at the same $k_x$ as the seed laser, is predominantly Rayleigh scattering from the seed and pump lasers, and indeed scales linearly with pump power.
Similarly the PL collected in region B corresponds to Rayleigh scattering from the input lasers.
Differently to A however, when the seed laser is absent or when it is positioned at $k_y=0$, both Rayleigh and parametric scattering overlap in B.
When the seed laser is positioned at any finite $k_y$ the PL in B becomes linear with pump power, and only region C shows direct evidence of the OPA process. 
We observe a similar trend in theory, see \cref{fig:4}b. 
There the nonlinear behaviour of C is even more evident due to the smaller reciprocal waist used in the simulations, \qty{0.001}{\mu m^{-1}} c.f. \qty{0.1}{\mu m^{-1}} in the experiment.

\subsection*{Robustness of the OPA}

\begin{figure}
    \centering
    \includegraphics[]{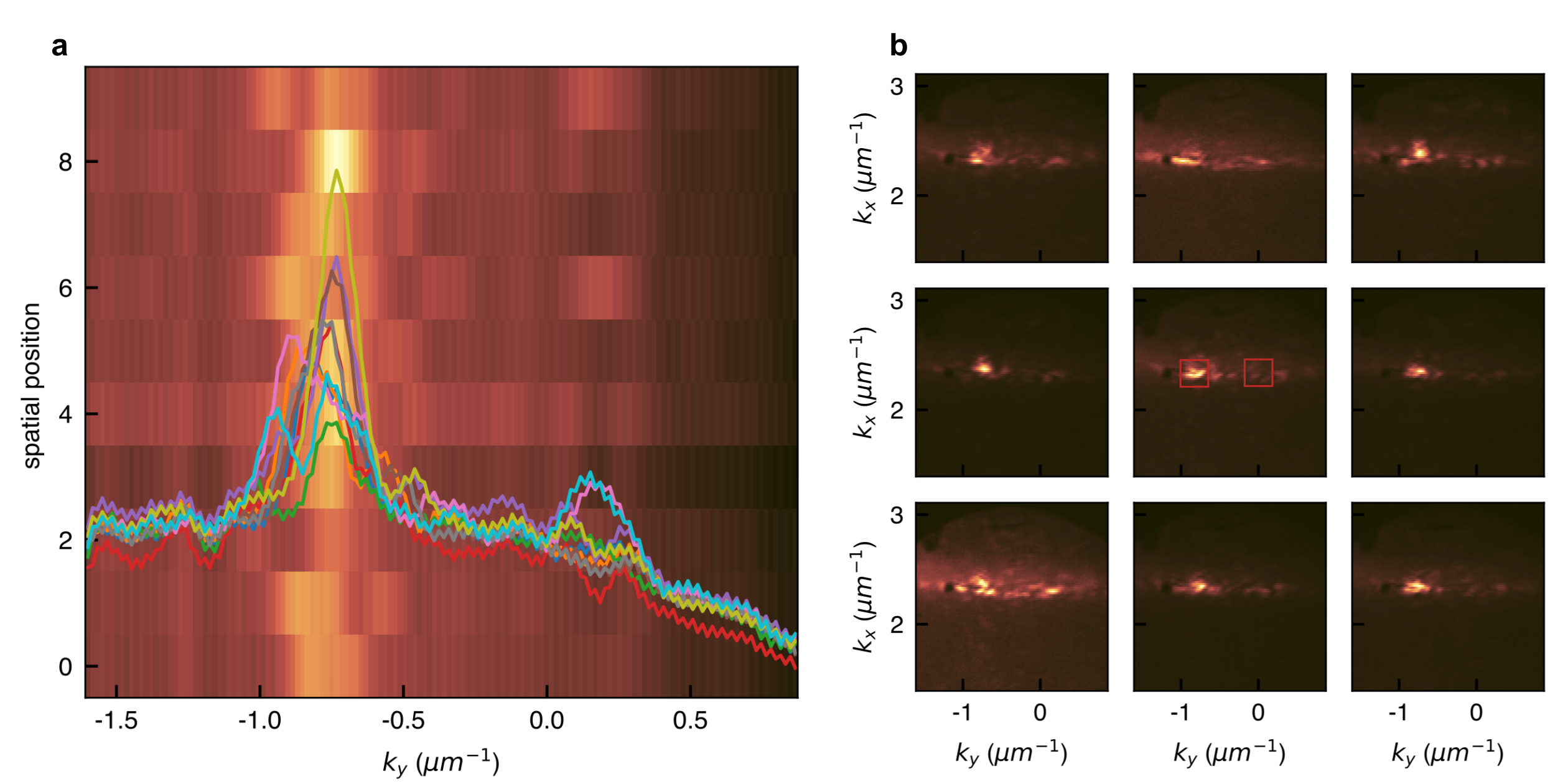}
    \caption{\textbf{Robustness of the OPA.} \textbf{a.} $k_x$ integrated reciprocal-space profiles for nine different spatial positions, demonstrating the phase-matching power and positional stability. \textbf{b.} Non integrated PL for 9 different positions in space showing that even though the average scattering in a given ROI might be similar, the PL distribution changes much more for the Rayleigh scattering than the phase-matched region. 
    }
    \label{fig:5}
\end{figure}

The OPA process generally depends on the local characteristics of the waveguide, such as the surface roughness and the presence of defects.
\Cref{fig:5} shows that moving the position of the waveguide, while keeping the laser configuration identical, the power and position of the idler are largely stable even if the distribution of the linear scattering changes significantly.

Shifting the waveguide along $x$ by a few tens of micrometers in 9 steps the collected PL distribution from the idler at $k_y=\qty{-0.8}{\mu m^{-1}}$ shows a clear localization and average power fluctuations of 18\%, see \cref{fig:5}a.
However, at $k_y=\qty{0}{\mu m^{-1}}$ where Rayleigh scattering dominates, the two-dimensional PL distribution shows no clear localization and is drastically affected by disorder; compare the PL in the marked ROI for all waveguide positions in \cref{fig:5}b.
Similarly, the position of maximum PL, indicating the directionality of the OPA, is at $k_y=\qty{-0.8(0.05)}{\mu m^{-1}}$.
The small fluctuations in $k_y$ are smaller than the reciprocal waist of the seed beam equal to \qty{0.1}{\mu m^{-1}}, indicating that slight inhomogeneity of the seed beam profile induced by surface defects might be responsible for fluctuations.
The seed beam position in reciprocal space, that could give rise to these fluctuations, is perfectly stable.
Considering typical pointing fluctuations in our experiment of a few micrometers, they translate to fluctuations in $k$ of the order of $10^{-3}$ for an imaging lens with a $NA=0.35$ and a clear aperture of \qty{5}{cm}; in any case the positional stability is unlikely to be changing significantly for the duration of this measurement.

\section*{Discussion}

In summary we demonstrated the feasibility of iso-energetic, OPA in a photonic crystal waveguide platform.
The topology of this hyperbolic metamaterial, protected by its C$_2$ symmetry, creates a BiC state at the lower branch, and a lossy state at the upper M-shaped branch.
The lossy state provides an ideal point of incidence for the pump laser.
The efficiency of the parametric process that ensues largely scales as $gN/\gamma$ where $gN$ is the chemical potential of the pump and $\gamma$ is the average loss of the signal and idler states. 
These correspond to the TE$_1$ modes of the waveguide whose losses are mainly photonic and not doubled as it happens for the lossy state in the vicinity of $\mathbf k=0$.

Although the parametric process is highly efficient it is less selective in energy and direction due to the small effective mass of the M-shaped branch around $\mathbf k=0$ and the preferential scattering direction imposed by the etched grating.
Using a seed laser we recover both energy and direction selectivity that now depend predominantly on the parameters of the seed laser.
We have shown the OPA process to be robust against surface defects, with its efficiency and directionality affected at the few percent level.

As opposed to microcavity platforms, where the strong coupling itself is used to create the phase-matching at the inflection point giving imbalanced OPO with a non-propagating signal beam at $\mathbf k=0$, the waveguide achieves this through dispersion engineering; strong-coupling is only used to induce the necessary nonlinearity in the system.
This additional flexibility allows to design different OPA processes at the same or different energies.
Iso-energetic process can be achieved in different platforms~\cite{abbarchi.PhysRevB.83.201310,ardizzone.scientific.reports2013}, however our waveguide offers some significant advantages.
In embedded microcavities the photon extraction efficiency is limited to the leakage rate of the Fabry-P\'erot resonator which necessarily low for high quality resonators and microwires show significant Rayleigh scattering due to their extreme aspect ratios.
The waveguide overcomes these limitations since photons are confined through total internal reflection and the etched grating has only subwavelength features which reduce scattering at the NIR optical wavelengths used here.

Optical parametric processes are ideally suited for the realization of all-optical microscopic devices~\cite{savasta.PhysRevLett.90.096403} where manybody effects and photon-photon correlations play an important role such as random number generators~\cite{marandi_all-optical_2012}.
Iso-energetic processes specifically are inherently better for the study of correlations and photon statistics ~\cite{ardizzone_bunching_2012}.
The dispersion engineering in our photonic crystal waveguide makes for a device with fast propagating photons in opposite directions that could be collected using separate coupling gratings with high efficiency and minimal crosstalk and as drastically simplify correlation studies.
At the same time, the OPA process can be considered an all-optical NOT gate~\cite{sannikov.natcomm2024}; a basic non-universal gate in computation.
Additional engineering can lead to cascadable gates with fast propagating photons making the photonic crystal waveguide an appealing optical computation platform.

These devices are simple to fabricate in GaAs heterostructures but can also be developed for room temperature devices using different materials, e.g. perovskites and two-dimensional transition metal dichalcogenides where parametric processes have already been observed~\cite{wu_perovskite_2021,zhao.nature.nanotechnology2022}.
The use of applied electric fields allows for more flexible engineering opportunities enabling different scattering channels and a polarisation degree of freedom~\cite{wang2024electricallygeneratedexcitonpolaritons}.

\section*{Methods}

\subsection*{Optical measurements}

The sample is held in Montana cryostation cold finger closed loop helium cryostat at approximately \qty{18}{K}.
The laser employed in this experiment is an ultra-narrow linewidth, continuous-wave M$^2$ Soltis laser. 
The excitation beam is split into two independent lines. 
The first line, referred to as the pump, excites the sample at $k=0$. 
The second excitation line, the seed, excites the same spatial position but at a finite angle to achieve resonance with the TE$_1$ mode through a 0.35 NA photographic objective.
To enable practical reciprocal space beam steering, the seed line incorporates a back-reflector mounted on a pair of translational stages. 
The power of the two beams is independently controlled by motorized waveplates and polarizers.

Given the resonant configuration of the experiment and the TE nature of the investigated modes, the excitation polarization is set to be diagonal, while cross-polarization is used in the detection to reduce the reflected excitation laser. 
This configuration effectively reduces the efficiency of both excitation and detection by a factor of 0.5.
The detection line is built in a 4f configuration, along which the collected photoluminescence is filtered in reciprocal space to remove the portions of the collected signal containing either the pump or the seed. 
Finally, the reciprocal space plane is reconstructed onto the entrance slits of a Horiba 550 monochromator, and the images are collected using a Hamamatsu Orca R$^2$ CCD camera.

The system is driven by an automated routine that collects reciprocal space emission data while scanning the injection power.
The data presented in the figures are post-processed as follows: For every dataset, three ROIs are defined along the TE$_1$ branch, and a fourth one is defined that does not contain any modes and is used to estimate the background counts. 
The total population within the four ROIs is integrated.

\subsection*{Fabrication}

The grating was fabricated by spinning ZEP520a 50\% resist at \qty{2000}{rpm} onto a \qty{510}{nm} thick GaAs/Al$_{0.4}$Ga$_{0.6}$As heterostructure waveguide. 
The resist was patterned using a Raith EBPG 5200 electron beam lithography tool and developed in amyl-acetate for 1 minute. 
Etching was performed with an Oxford ICP-Chlorine etcher to achieve a total depth of \qty{170}{nm}. 
The resist was then removed using dichloromethane, followed by the conformal deposition of a \qty{10}{nm} aluminum oxide layer through atomic layer deposition.

\subsection*{System Hamiltonian}

The hamiltonian of the photonic crystal waveguide in reciprocal space $\hat H(\mathbf k)$ comprises four coupled elliptical cones whose centres are shifted in $k_x$, $C(\mathbf k) = \sqrt{(k_x - k_{x,0})^2 + k_y^2}$.
Each cone describes a co- or counter-propagating TE$_{\pm 0, \pm 1}$ guided mode.
The grating folds the modes along $k_x$ and induces diffractive energy couplings $U_{00}$, $U_{11}$, and $U_{01}$.
All modes are coupled to the excitonic cloud at energy $\hbar\omega_x$ with Rabi couplings $\Omega_0$ and $\Omega_1$.
Including the dissipation matrix, where $\gamma$, $\gamma_x$ are the photonic and excitonic losses respectively, we get the following $5\times 5$ hamiltonian

\begin{equation} \label{eq.H}
\hat{H}(\mathbf{k}) = \hat{H}_0(\mathbf{k}) + i\hat{H}_\gamma(\mathbf{k}) = \begin{pmatrix}
    C_{+0}(\mathbf k) & U_{00} & 0 & U_{01} & \Omega_0 \\
    U_{00} & C_{-0}(\mathbf k) & U_{01} & 0 & \Omega_0 \\
    0 & U_{01} & C_{+1}(\mathbf k) & U_{11} & \Omega_1 \\
    U_{01} & 0 & U_{11} & C_{-1}(\mathbf k) & \Omega_1 \\
    \Omega_0 & \Omega_0 & \Omega_1 & \Omega_1 & \omega_x
\end{pmatrix} - 
i\begin{pmatrix}
    \gamma & \gamma & 0 & \gamma & 0 \\
    \gamma & \gamma & \gamma & 0 & 0 \\
    0 & \gamma & \gamma & \gamma & 0 \\
    \gamma & 0 & \gamma & \gamma & 0 \\
    0 & 0 & 0 & 0 & \gamma_x
\end{pmatrix}.
\end{equation}

\subsection*{Numerics}

One of the achievements of the experiment is to choose the direction of propagation of the modes generated by the nonlinear scattering (OPO). 
The direction is fixed by using a seed laser at a given wavevector in the two-dimensional dispersion. 
In numerics, we get a sufficient understanding of this OPA phenomenon with a one-dimensional dispersion -- it is achieved by turning on and off the parameters that cause the different types of scattering in the system.

The dispersion of the physical model is modelled by \cref{eq.H}, which describes the coupling of four photonic mode to an excitonic mode. 
We define $\Psi$ as the vector containing the wavefunctions of all these modes, such that $\Psi=\left(\psi_{+0}, \psi_{-0}, \psi_{+1}, \psi_{-1}, \psi_{x}\right)^T$. 
The dynamics of these wavefunctions are described by the Gross-Pitaevskii equation, which at the mean-field level and in reciprocal space is ($\hbar=1$)
\begin{equation} \label{eq.GPE}
i \partial_t \Psi(\mathbf{k}) = \hat{H}(\mathbf{k})\Psi(\mathbf{k}) + \sum_{ph} \sum_{\mathbf{k}'} V_d(\mathbf{k}-\mathbf{k}') \psi_{ph}(\mathbf{k}') + g_x \sum_{\mathbf{k}_1,\mathbf{k}_2} \psi^*_x(\mathbf{k}_1+\mathbf{k}_2-\mathbf{k})\psi_x(\mathbf{k}_1)\psi_x(\mathbf{k}_2) + \left(1, 1, 1, 1, 0\right)^T \left[F_p(\mathbf{k},t) + F_s(\mathbf{k},t)\right]. 
\end{equation}
The first term on the right hand side sets the dispersion of the system through the Hamiltonian in \cref{eq.H}. 
The following term is a static disorder field, $V_d$, acting on the photonic modes. 
Disorder is required to scatter polaritons between different momentum states and, in this way, to model resonant Rayleigh scattering~\cite{PhysRevB.76.115326,PhysRevB.82.081301,PhysRevB.92.035307}. 
The next term is the nonlinear interaction of the excitonic mode, whose strength is governed by $g_x$, and which describes the scattering processes involved in the OPO regime. 
Finally, the last term contains the pump, $F_p$, and the seed, $F_s$, lasers coherently driving the four photonic modes. 

The pump and seed lasers are modelled with the following Gaussian profiles in reciprocal space
\begin{equation}\label{eq.pump}
    F_{p,s}(\mathbf{k},t) =  \frac{F_{p,s}}{
    (w\sqrt{\pi})^{\frac{1}{2}}
    }
e^{-\frac{(\mathbf{k}-\mathbf{k}_{p,s})^2}{2w^2}} e^{-i \omega_{p,s} t}.
\end{equation}
The pump and seed spots are centred at momenta $\mathbf{k}_{p}$ and $\mathbf{k}_{s}$, have energy $\omega_{p}$ and $\omega_s$, and strength given by $F_p$ and $F_s$. 
Their width is assumed to be the same and regulated with the parameter $w$.

The aim of our simulations is to verify some of the experimental observations. 
We solve the coupled differential equations of \cref{eq.GPE} in one dimension and starting from a noisy initial state. 
The pump laser drives the middle of the lossy branch, at $k_p=0$, which is at the crossing of the two counterpropagating TE$_0$ modes. 
In this case OPO scattering can happen iso-energetically with the signal and idler populating the two TE$_1$ branches of the same energy band. 
The same modes can also be populated by linear Rayleigh scattering. 
In order to model the OPA, we add a seed laser at the position of the signal mode. 
Finally, the population of the mode of interest, i.e., the idler, is characterized by integrating the density of the corresponding TE$_1$ mode over a region of interest in phase space: a small window around the idler mode.

In particular, to emulate the two-dimensional behaviour with one-dimensional simulations, we consider the effect of both the nonlinear coupling $g_x$ and the couplings between TE$_0$ and TE$_1$ modes, which we set to be $U\equiv U_{00}=U_{11}$ and $U_{01}=U/2$ here. 
By setting $g_x$ and/or $U$ to 0, we can turn the effects of nonlinear and linear scattering off, respectively. 
Then, we perform scans of the power of the pump and seed beams, as it is done in the experiment, and measure the population of the idler mode.

The dependence in pump power from the ROIs in \cref{fig:4} of the main text can be emulated as follows:
When $g_x=0$ and $U=0$, nothing should happen irrespectively of using a seed beam or not.
When $g_x=0$ and $U\neq 0$, only linear scattering is possible. 
Using the seed does not affect the scattered PL much except for the addition of scattered PL. 
This emulates looking at the region B when OPA happens diagonally.
When $g_x\neq 0$ and $U=0$, there is no linear scattering but only parametric. 
Using a seed triggers OPA and lowers the threshold. 
This emulates region C when there is a seed at the limiting case of small linear coupling.
When $g_x\neq 0$ and $U\neq0$, both linear and nonlinear scattering occur. 
This emulates the region B when there is the pump only or region C with seed on.

Values of parameters in the simulations:
The experimental dispersion of the lower polariton modes is reproduced with the following parameters: 
$\omega_x=\qty{-3.3}{meV}$,
$U=\qty{7.5}{meV}$, 
$\Omega_0=\Omega_1=\qty{8}{meV}$. 
The dissipation is $\gamma = \qty{0.51}{meV}$, $\gamma_x = 0.001\gamma$. 
The off-diagonal $\gamma$'s in \cref{eq.H} are set to $90\%$ of the diagonal $\gamma$'s value in order to keep the numerics stable.
The photonic modes have a slope of $\pm \qty{37}{meV \mu m}$, the TE$_0$ pass through the origin of the dispersion, and the distance between TE$_0$ and TE$_1$ modes in reciprocal space is \qty{2.45}{\mu m^{-1}}.
Spatial disorder follows a normal distribution with magnitude $V_d=\qty{0.001}{meV}$.
The dispersion with $U=0$ is not physical, but parameters are chosen such that the second band matches experiments well: all parameters are the same, except for the photonic modes: TE$_0$ modes are offset by $\qty{-6}{meV}$ from the origin, and the distance between TE$_0$ and TE$_1$ modes in reciprocal space is set to \qty{2.6}{\mu m^{-1}}. %

\bibliography{bibliography}

\section*{Acknowledgements}

We thank Paolo Cazzato for technical support.
This project was funded by PNRR MUR project: `National Quantum Science and Technology Institute' - NQSTI (PE0000023);
PNRR MUR project: ‘Integrated Infrastructure Initiative in Photonic and Quantum Sciences’ - I-PHOQS (IR0000016);
Quantum Optical Networks based on Exciton-polaritons - (Q-ONE) funding from the HORIZON-EIC-2022-PATHFINDER CHALLENGES EU programme under grant agreement No. 101115575;
Neuromorphic Polariton Accelerator - (PolArt) funding from the Horizon-EIC-2023-Pathfinder Open EU programme under grant agreement No.  101130304;
the project ``Hardware implementation of a polariton neural network for neuromorphic computing'' – Joint Bilateral Agreement CNR-RFBR (Russian Foundation for Basic Research) – Triennal Program 2021–2023;
the MAECI project ``Novel photonic platform for neuromorphic computing'', Joint Bilateral Project Italia - Polonia 2022-2023;  
the PRIN project ``QNoRM: A quantum neuromorphic recognition machine of quantum states'' - (grant 20229J8Z4P).
Views and opinions expressed are however those of the author(s) only and do not necessarily reflect those of the European Union or European Innovation Council and SMEs Executive Agency (EISMEA). Neither the European Union nor the granting authority can be held responsible for them.

\section*{Author contributions statement}

AG conducted the experiments and together with DT analysed the results.
DW, PC and MS worked on the theoretical and numerical framework.
DS supervised the work.
DB and MDG assisted with the experimental setup and interpretation of the data.
All authors contributed to discussions and editing of the manuscript.

\section*{Data availability statement}

The data that support the findings of this study are available from the corresponding author upon reasonable request.

\section*{Additional information}

\textbf{Competing interests}.
The authors declare no competing interests.

\clearpage

\end{document}